\documentclass[preprint,showpacs,preprintnumbers,amsmath,amssymb,showkeys]{revtex4}

\usepackage{graphicx}
\usepackage{dcolumn}
\usepackage{bm}
\newcommand{\dir}{graphs/}
\renewcommand{\vec}[1]{{\bf #1}}

\begin{document}


\title{Density Matrix Renormalization for model reduction in nonlinear dynamics}

\author{Thorsten Bogner}
\homepage{http://www.physik.uni-bielefeld.de/theory/cm/}
\affiliation{
Condensed Matter Theory Group, Fakult\"at f\"ur Physik, Universit\"at Bielefeld
}
\date{April 26, 2007}

\begin{abstract}
We present a novel approach for model reduction of nonlinear dynamical systems
based on proper orthogonal decomposition (POD). Our method, derived from
Density Matrix Renormalization Group (DMRG), provides a significant reduction
in computational effort for the calculation of the reduced system, compared to
a POD. The efficiency of the algorithm is tested on the one dimensional
Burgers equations and a one dimensional equation of the Fisher type as
nonlinear model systems.
\end{abstract}


\maketitle
Nonlinear dynamical systems arise in many fields as physics, e.g.
turbulence~\cite{sirovich_1987}, mathematics and biology~\cite{murray}.
They often require a
significant high number of degrees of freedom(dof) for simulation with
suitable accuracy. 
For a large class of systems the solutions are regular,
nevertheless the influence of the nonlinearity is essential.
Here model reduction (MR) can lead to an efficient
description, if the dynamics is effectively confined to a lower dimensional
attractor in phase space.

The aim of this work is to develop an algorithm that can find a reduced
model for a given system, usually derived from a partial differential
equation.

One method to obtain such a reduced description is the so called proper
orthogonal decomposition~\cite{sirovich_1987, lucia} (POD).
It is obtained by
calculating the eigenvectors of the spatial covariance matrix of the field
over the phase space, typically by using a empirical spatial covariance matrix
from a number of realization of the dynamic evolution.
The method itself is linear in that the phase space is reduced to a subspace
(in which the relevant Attractor has to be embedded). Nevertheless it accounts
for the nonlinearity and gives the 'optimal' linear reduction possible.
By definition the POD requires a simulation of the full, unreduced system
and a diagonalization of a symmetric matrix of similar size.
While the later can be circumvented by the method of snapshots of
Sirovich~\cite{sirovich_1987}, the simulation in unavoidable.

Within our approach, we try to calculate 'approximate' POD modes
without simulating the full, unreduced system. This is achieved by following
concepts from density matrix renormalization. Effectively, we
adapt a system of already low dimensionality,
to reproduce the full dynamics optimally. Consequently, all calculations
are performed on low dimensional systems. In exchange several calculation
steps have to be performed, but their number is proportional to the size
of the full system of interest. Practically this is one possible way
to study large dynamical systems, although within this work we are still
restricted to spatially one dimensional systems.
Beside from possible benefits for the efficiency, an interesting question is
whether it is possible to reconstruct dynamic behavior of a system
from the knowledge of subsystems only.

This paper is organized as follows. First, we introduce the formulation of the
equations defining the dynamical system. This includes the use of higher order
tensors to describe the nonlinear part of the generator of the time evolution.
Further, the discretization of the three model equations, namely the linear
diffusion equation, the Burgers equation and a nonlinear diffusion equation,
is presented.
Then the type of orthogonal projection, which is used in this work, is
introduced and the basic concept of the proper orthogonal decomposition is
recapitulated. After a brief outline of the single particle DMRG approach, the
method devised in this paper is presented.
The numerical results, including a comparison of our method with standard
techniques, are given consecutively. This is followed by a short discussion
of our approach and the conclusions. The appendix finally contains an analysis
of the optimal reduction for the linear case.

\section{The Problem}\label{problem}
\subsection{The Dynamical System}
We consider discretized versions of nonlinear evolution
equations of the form
\begin{eqnarray}\label{evo1}
\partial_t\Phi_i =  (G(\Phi)\Phi)_i\nonumber\\
=L_{ij}\Phi_j + Q_{ijk}\Phi_j\Phi_k 
+ K_{ijkl}\Phi_j\Phi_k\Phi_l \label{system}
\end{eqnarray}
Here $\Phi$ is the field, $G(\Phi)$ is the nonlinear generator of evolution
and we make use of the sum convention.
The contributions $L$, $Q$ and $K$ are the
linear, the quadratic and the cubic
part, respectively, of the generator of evolution.
Higher order terms can also be considered, but the number of nonlinear terms
should be finite for our approach. Note that $Q$ and $K$ are third and fourth
order tensors and have the corresponding transformation properties.
The dynamical system described in Eq.(\ref{system}) is typically derived from
a partial differential equation (PDE).
The spatial discretization is then done by finite differences 
or equivalently by linear finite elements(FE). The temporal discretization
is done by the simple explicit Euler method, although this choice is not
relevant for our method.
Here, we restrict ourselves to the spatially one dimensional case.
In the following we exemplify our approach on simple toy problems.

\subsection{The Linear Diffusion Equation}
The diffusion equation describes diffusive transport of a scalar field, e.g.
heat transport, in a medium. For homogenous media it is given by
\begin{equation}\label{diffu1}
\frac{\partial}{\partial t}\Phi(x,t) = d\Delta\Phi(x,t)\quad x\in[0,1]
\end{equation}
with the diffusion constant $d$. Spatial discretization of the interval
$[0,1]$ with $N$ nodes gives for the discrete Laplace operator with
second order accuracy in $\Delta x$ the following $N\times N$ matrix
\begin{equation}\label{laplace}
\Delta_N=\frac{1}{\Delta x^2}\left(
\begin{array}{ccccc}
-1& 1 &      &      &  \\
1 &-2 & 1    &      &  \\
  & 1 &\ddots&\ddots&  \\
  &   &\ddots&  -2  & 1\\
  &   &      &   1  &-1
\end{array}
\right)
\end{equation}
Here homogenous Neumann conditions are assumed for $x=0$ and $x=1$,
the spatial discretization step size is $\Delta x = \frac{1}{N}$.
The explicit Euler method gives for the discrete time evolution with
time step size $h_t$ the following equation
\begin{equation}\label{diffu_euler}
  \tilde{\Phi}(\tilde{x},t_{n+1}) = \tilde{\Phi}(\tilde{x},t_{n})
  + dh_t\Delta_N\tilde{\Phi}(\tilde{x,t_{n}})
\end{equation}
where $\tilde{\Phi}$ and $\tilde{x}$ are $N$-dimensional vectors, indicated by
$\tilde{\cdot}$. Thus the linear
part $L$ in Eq.(\ref{system}) is given by 
\begin{equation}\label{eq:diffuL}
L=dh_t\Delta_N.
\end{equation}
The nonlinear contributions in Eq.(\ref{system}) vanish for the linear
diffusion equation.
\subsection{The Burgers Equation}
As one nonlinear example we consider the Burgers equation~\cite{burgers}.
It describes a diffusive as well as a convective transport of a scalar
field $\Phi$ and is given by
\begin{equation}\label{burgers1}
\frac{\partial}{\partial t}\Phi = d\Delta\Phi + \nu (\Phi\nabla)\Phi.
\end{equation}
This equation is similar to the linear diffusion equation Eq.(\ref{diffu1})
but with an additional term $\nu (\Phi\nabla)\Phi$, describing the convection.
This term is quadratic in the field $\Phi$ and can be discretized in the form
of $Q$ in Eq.(\ref{system}). For one space dimension, the $\nabla$ operator
is simply the spatial derivative. This can be discretized with second order
accuracy in $\Delta x$ as~\cite{numerical_recipes}
\begin{equation}\label{derivative}
D_{x,N}=\frac{1}{\Delta x}\left(
\begin{array}{ccccc}
-1& 1 &      &      &  \\
-1& 0 & 1    &      &  \\
  &-1 &\ddots&\ddots&  \\
  &   &\ddots&  0   & 1\\
  &   &      &  -1  & 1
\end{array}
\right)
\end{equation}
The term $(\Phi\nabla)$ is also known as convective derivative. In 1D
the discretization is given by multiplying the rows of $D_{x,N}$ with
the components of $\tilde{\Phi}$:
\begin{equation}\label{conv_derivative}
(\Phi\nabla)_{N, i,j}= \tilde{\Phi}_i D_{x,N, i,j}
\end{equation}
here $i,j$ indicate the component of the matrix/vector. Choosing
\begin{equation}\label{eq:burgersQ}
Q_{i,j,k}:=\nu D_{x,N, j,k}\delta_{ij}
\end{equation}
gives a discretization of the convection term,
as defined in Eq.(\ref{conv_derivative})
\begin{eqnarray}\label{conv_derivative2}
\sum_{j,k}Q_{i,j,k}\tilde{\Phi_{j}}\tilde{\Phi_{k}} =\nonumber
\nu \sum_{j,k}D_{x,N, j,k}\delta_{ij}\tilde{\Phi_{j}}\tilde{\Phi_{k}}\\ =
\nu \sum_{k}\tilde{\Phi_{i}}D_{x,N, i,k}\tilde{\Phi_{k}} =
\nu \left(\Phi\nabla \right)_{N}\tilde{\Phi}.
\end{eqnarray}
\subsection{Nonlinear Diffusion}
We consider here a Diffusion equation with a nonlinearity that resembles the
action-potential part of the one dimensional
FitzHugh-Nagumo(FN)~\cite{fitzhugh_1961, yanagita_2005} equation.
In particular the dynamics is defined by
\begin{equation}\label{FN1}
\frac{\partial}{\partial t} \Phi = \Delta\Phi - \Phi(1-\Phi)(a-\Phi)
\end{equation}
where $a$ is a constant. Eq.(\ref{FN1}) has stable equilibria at
$\Phi\equiv 0$ and $\Phi\equiv 1$ and an instable equilibrium
at $\Phi\equiv a$.
The nonlinear term is cubic in the field. It can be rewritten as
$-\Phi(1-\Phi)(a-\Phi)=-\Phi^3+(1+a)\Phi^2-a\Phi$. Here the powers of $\Phi$
are defined component wise. The cubic part $-\Phi^3$, e.g. is discretized by
\begin{equation}\label{eq:FHK}
K_{i,j,k,l}=-\delta_{ij}\delta_{ik}\delta_{il}
\end{equation}
since
\begin{equation}\label{cubic1}
\sum_{j,k,l}\left(-\delta_{ij}\delta_{ik}\delta_{il}\right)
\tilde{\Phi_j}\tilde{\Phi_k}\tilde{\Phi_l} =-\tilde{\Phi_i}^3
\end{equation}
Similarly, the quadratic part becomes
\begin{equation}\label{eq:FHQ}
Q_{i,j,k}=\left(1+a\right)\delta_{ij}\delta_{ik}
\end{equation}
and the linear part together with the contribution from the diffusive
term is
\begin{equation}\label{eq:FHL}
L_{i,j}=dh_t\Delta_{N,i,j} - a\delta_{ij}.
\end{equation}
Since we deal only with discretized fields $\tilde{\Phi}$ in the following,
we drop the notation $\tilde{\cdot}$ for discrete variables.

\subsection{The Reduction}
To obtain a reduced model, we will project the phase space to a lower
dimensional subspace. Thus the reduction is linear, which simplifies the
problem significantly. For nonlinear reduction see
e.g.~\cite{gorban_karlin_zinovyev_2004, steindl_troger_2001}.
For a linear projection we only need a basis of the relevant subspace, e.g
given by the column vectors of a $N\times M$ matrix $B$, where $N$ is the
dimensionality of the phase space and $M$ that of the subspace ($N>M$).
We will always assume an orthonormal basis in the following since $B$ can
always be brought to this form.
This basis spans the range of the projection operator $P$, which is
defined by
\begin{equation}\label{eq:P_def}
P=BB^\dag.
\end{equation}
The reduced dynamics is given by
\begin{eqnarray}\label{evo2}
\partial_tP\Phi=PLP\Phi + P\left(QP\Phi P\Phi\right)
+ P\left(KP\Phi P\Phi P\Phi\right)\label{system_red}
\end{eqnarray}
One can write this equation directly for the reduced phase space which is
only $M$ dimensional by using
\begin{eqnarray}\label{red_var1}
\hat{\Phi}=B^\dag\Phi\\
\hat{L}=B^\dag LB\label{eff_L}\\
\hat{Q}_{i,j,k}=\sum_{a,b,c}B_{i,a}^\dag Q_{a,b,c}B_{b,j} B_{c,k}\label{eff_Q}\\
\hat{K}_{i,j,k,l}=\sum_{a,b,c,d}B_{i,a}^\dag K_{a,b,c,d}B_{b,j} B_{c,k} B_{d,l}\label{eff_K}
\end{eqnarray}
This gives the reduced equations which are still in
the form of Eq.(\ref{system}) as
\begin{eqnarray}
\partial_t\hat{\Phi} =  \hat{L}_{ij}\hat{\Phi}_j +
\hat{Q}_{ijk}\hat{\Phi}_j\hat{\Phi}_k 
+ \hat{K}_{ijkl}\hat{\Phi}_j\hat{\Phi}_k\hat{\Phi}_l
\label{system_red2}
\end{eqnarray}
While this dynamics is defined on a smaller, $M$ dimensional phase space
it has to be noted that the operators are now typically dense, i.e. most
entries in the tensors $L$, $Q$ and $K$ are nonzero.
To define the reduction we have to make a choice for the relevant degrees
of freedom that span the range of $P$, i.e. the orthonormal basis (ONB) $B$.
Natural criteria for the determination
of $B$ would be based on the difference
\begin{equation}\label{eq:diff_err}
{\bf{E}(t)}=\Phi(t)-BB^\dag\hat{\Phi}(t) = (\openone-P)\Phi(t),
\end{equation}
e.g. the $L^2$~-error
\begin{equation}\label{eq:diff_err2}
E(t):=||\bf{E}(t)||_2.
\end{equation}

For linear systems, as e.g. the linear diffusion equation, it can be
shown (see section~\ref{Appendix})
that the projector onto the eigenstates with lowest absolute eigenvalue
of the generator of the time evolution (i.e. $L$ in Eq.(\ref{evo1}))
lead to a minimal $L^2$-error for long and short enough times.
In the long time limit this approximation becomes even arbitrary accurate
as long as the discarded eigenvalues are smaller than zero.
This can be extended to nonlinear systems using the Proper Orthogonal  
Decomposition(POD).
\begin{figure}
\includegraphics[width=0.42\textwidth]{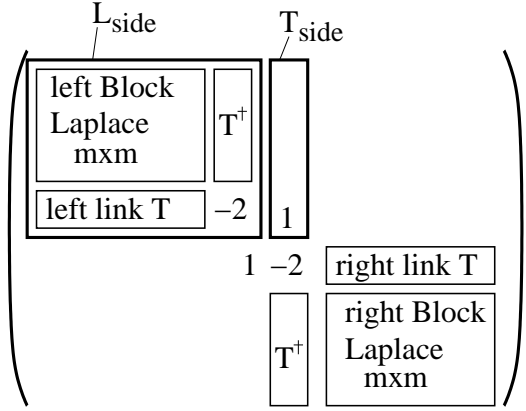}
\caption{Assembly of the superblock Laplace operator}
\label{fig:blocking1}
\end{figure}
\begin{figure}
\includegraphics[width=0.40\textwidth]{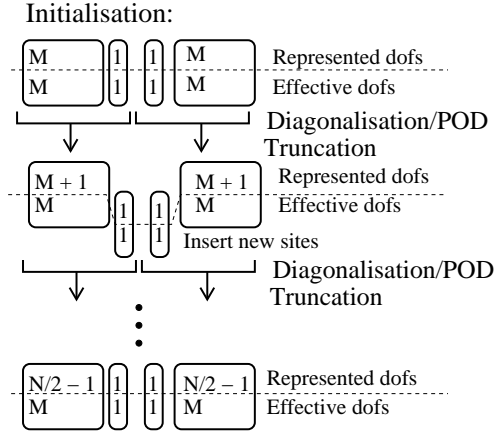}
\caption[Graphical illustration of the DMRG initialization (or warmup) scheme]
    {Graphical illustration of the DMRG initialization (or warmup) scheme}
\label{fig:dmrg_init}
\end{figure}
\begin{figure}
\includegraphics[width=0.42\textwidth]{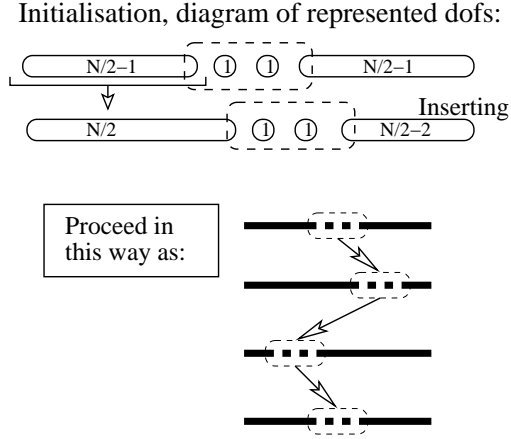}
\caption[Graphical illustration of the DMRG iteration (or sweeping) scheme]
{Graphical illustration of the DMRG iteration (or sweeping) scheme}
\label{fig:dmrg_iter}
\end{figure}

\section{Proper Orthogonal Decomposition}
The proper orthogonal decomposition is a linear projection method which
is widely used in model reduction. On this topic an extensive literature
exist. Some examples are~\cite{lorenz_1956, sirovich_1987, berkooz_holmes_lumley_1996,
  rowley_colonius_murray_2003, noack_afanasiev_morzynski_tadmore_thiele_03}.
A short explanation of POD together with the method of snapshots is also given
in~\cite{damodaran_willcox_2003}.
One of the advantages of this method is the possibility to
incorporate information from the nonlinear dynamics to obtain
a linear reduction.
The basic idea is to generate sample trajectories by simulating the
dynamical system of interest. Then the spatial
two point correlation matrix $C$ is calculated
\begin{equation}
\label{correl}
C_{i,j}:=\left<\Phi(x_i,t)\Phi(x_j,t)\right>_{\mathbf{T}}
\end{equation}
Here $\left<\right>_{\mathbf{T}}$ denotes an average over all
sample trajectories. 
The eigenvectors of this symmetric matrix, which correspond to the highest
eigenvalues span an 'optimal' subspace in the sense
that the average least square truncation error
\begin{equation}
\label{epsilon}
\epsilon:=\left<||\Phi(x,t)-P\Phi(x,t)||^2\right>_{\mathbf{T}}
\end{equation}
is minimal, see e.g.~\cite{sirovich_1987, antoulas}.
The (orthonormal) basis vectors of this subspace constitute the columns of
the matrix $B$ which defines the projection operator $P$,
see Eq.(\ref{eq:P_def}).
The practical application in the following algorithm is simple: Once the
operators $L$, $Q$, and $K$ are calculated
and the initial conditions are given,
we can simulate the dynamics of the field $\Phi$, e.g. by
Eq.(\ref{system}). For the reduced system Eq.(\ref{system_red2}) is used
instead.
During the simulation the data for the covariance
matrix $C$ is accumulated, if necessary several simulation runs are performed
using different initial condition with appropriate weighting.
The eigenvectors of $C$ are calculated using standard
methods~\cite{numerical_recipes}. Then $B$ is constructed from those
eigenvectors corresponding to the highest eigenvalues. The discarded
eigenvalues can also provide information on the quality of the reduction.
The POD can be applied relatively independent from the actual
blocking method.
It should be noted that such a reduction is
optimal for describing the generated
dataset, but not necessarily optimal in reproducing the
underlying dynamics~\cite{rowley_2005}.

\section{Blocking Method}
Blocking methods were considered already earlier, e.g.~\cite{lucia_2003},
mainly 
because in many problems not all spatial regions are of similar interest.
Our motivation is different, we aim to decompose a calculation into
more feasible sub-problems.
In the linear case the basis $B$ can be calculated in principle
by simply diagonalizing $L$. Technically this is the same problem as
determining the eigenstates of the Laplace operator, which also describes
the single quantum mechanical 1D-particle in a
box. S.White~\cite{white_prl_68,white_prl_69,white_1993} used this toy model
for introducing 
the so called Density Matrix Renormalization Group (DMRG). This approach has
been since then applied most successfully to quantum many-body problems.
In the following we want to carry this analogy further. Instead of 
approximating the eigenvalues/states of a linear operator we use
a similar algorithm to obtain an approximate POD of a nonlinear system.
To this end a few modifications are necessary,
so we first summarize the original DMRG method.

\subsection{Single Particle DMRG}
In the DMRG toy problem the system is split into blocks of size $m$.
For each block
a block-Laplace operator is stored, as well as the links $T$ that define the
interaction with the neighboring sites. The assembly of the superblock
Laplace operator is pictorially presented in
Fig.~\ref{fig:blocking1}.
This superblock operator is a $(2m+2)\times(2m+2)$ matrix and
has to be diagonalized. From the eigenstates the so called target states
$\phi^i$, $i=1\ldots (m+1)$ are
selected, usually the low lying spectrum.
Since we have in this special case a single particle problem, the block
truncation matrix $R$ can be calculated simply by applying a Gram-Schmidt
orthogonalization to the block part of the target states.
\begin{equation}
\label{eq:Trunc}
R_{i,j}=\mbox{Gram-Schmidt}(\phi^i_{j}) \quad i=1\ldots(M+1), j\in\mbox{Block}
\end{equation}
For more general DMRG applications $R$ would be calculated by diagonalizing the
density matrix of the target state. $R$ is a $(m+1)\times m$ matrix which
projects the phase space of one system side to an effective block.

The effective block Laplacian $L_{eff}$ and the effective block link $T_{eff}$
are derived as follows
\begin{equation}
\label{reduceblock}
L_{eff} = R^\dag L_{side}R\qquad
T_{eff} = R^\dag T_{side}
\end{equation}
where the form of $L_{side}$ and $T_{side}$ are
depicted in Fig.~\ref{fig:blocking1} for the
left side. By this process effective blocks are calculated, that describe
a higher number of sites, but have numerically still $m$ degrees of freedom.

In DMRG this 'growing' of blocks is first used in an initialization step
until the superblock describes a sufficiently large system, see
Fig.~\ref{fig:dmrg_init}. Then an iteration
is carried out to increase accuracy. Here only one side is grown, while
the other is replaced by an already calculated block so that the effective
size of the superblock is constant, see Fig~\ref{fig:dmrg_iter}.

\subsection{DMRG-POD}
Basically three modifications are necessary to obtain a DMRG-POD algorithm
\begin{description}
\item[First,] instead of a diagonalization of the superblock operator, a
  POD on the superblock system has to be performed. This is composed of
  first, a simulation of the superblock system, 
  as defined in Eq.(\ref{system_red2}).
  Then the superblock correlation matrix from the generated data
  has to be diagonalized. This gives an orthonormal set of vectors
  which are the target vectors in the context of DMRG.
\item[Second,] to each sub-block there exist not only a linear sub-block
  operator but also higher order operators, given by third and higher order
  tensors, see Eq.(\ref{evo2}). These have to be updated in a similar way.
\item[Third,] for the POD the initial states for the sample trajectories are
  crucial. The initial states are defined for the full system. They have to be
  projected onto the superblock system which requires all truncation matrices
  explicitly.
\end{description}
Concerning the first point, this is no great difference, since the POD
(simulation and diagonalization of the correlation matrix)
returns also a orthonormal set of 'relevant' states (the POD modes) that serve
as target states, as described above.

Beside the linear operator ($L$ in Eq.(\ref{system})) which is assembled
identically as the superblock operator in DMRG, the higher order operators
have to be a assembled also. This is principally possible, but complex.
Here we use a simple trick. For all our models it is sufficient to know
the component-wise squaring operator $\Omega_{i,j,k}:=\delta_{ij}\delta_{ik}$.
(And in some cases the derivative operator which is linear and is
also assembled like the superblock Laplace operator.) $\Omega$
is purely diagonal,
so no links have to be stored and assembled. The reduction with
a truncation matrix $R$ is straightforward:
\begin{equation}
\label{reduceSQ}
\hat{\Omega}_{i,j,k} = \sum_{a,b,c}R_{i,a}\Omega_{a,b,c}R_{b,j}R_{c,k}
\end{equation}
From this the higher order tensors can be calculated directly, e.g.
for the Burgers equation
\begin{eqnarray}
\hat{Q}_{\mbox{Burgers},i,j,k}:=\nu
\sum_{l}\delta_{ij}\delta_{lj}\hat{D}_{x,N, l,k}\nonumber\\
= \sum_{l} \hat{\Omega}_{j,i,l}\hat{D}_{x,N, l,k}.\label{reduce_burgers}
\end{eqnarray}
For fourth and higher order operators this procedure is a bit memory
consuming. E.g for calculating $\Phi^3$ it is more efficient to
calculate first $\Phi_{\mbox{\scriptsize tmp}}:=\hat{\Phi}^2=\Omega\Phi\Phi$ and then
$\hat{\Phi}^3=\hat{\Phi}^2\hat{\Phi}=\Omega\Phi_{\mbox{\scriptsize tmp}}\Phi$.

The third point is a small disadvantage, since the projection operators
have all to be stored, which is not necessary in DMRG if only the energy
values are of interest. However, here as well as in DMRG it is possible
to expand a superblock state to a state of the original system as well as
project down a system state  to the superblock if all truncation matrices
are stored. The down-projection of the $N$-dimensional state is in
particular done by iteratively contracting the $m+1$ outermost sites of
e.g. $\Phi$ with
the corresponding block truncation matrix $R$. Apart from the memory
requirement this is simply a book keeping problem.

It should be noted that only $m+1$ most relevant states from the POD are used
as target states. Thus
only $m+1$ relevant states of the superblock are optimized although
it represents $2m+2$ dofs. This has to be considered in comparing the
results. However, the DMRG POD is nevertheless faster than the full POD,
see section~\ref{comp_load}.

To summarize: Apart from the POD itself, which is a standard technique, no
fundamental changes have to be implemented to get a DMRG-POD method from the
simple toy model DMRG. The assembly of linear operators has to be performed
in any case, only our method requires several operators. The assembly of the
$\Omega$ operator is even more simple, since all links vanish.
Reconstruction of full system states is also possible in DMRG, only it
is mandatory for our method for evaluating the correct initial conditions.
\section{Application}
For all applications we choose a finite differencing scheme
of second order accuracy, homogenous Neumann conditions at the boundaries
and the explicit Euler method for the calculations. The details are given in
section~\ref{problem}, above.
The boundary conditions
as well as the time integration method can be
chosen - more or less - arbitrarily. However, higher order finite
elements in the spatial discretization lead to additional interactions between
single dofs, i.e. a form of non-locality and do thereby complicate the
problem. For the reduced system size always four dofs were retained.
This is mainly for convenience and easy comparison. The success of the
method does not depend strongly on this choice.

As explained above, we measure the quality of a reduction by the $L^2$-error,
see Eq.(\ref{eq:diff_err2}). 
It has the same units as the fields $\Phi$ which are not further
specified. The time units are also arbitrary.
The error calculations in the following are performed in a separate program
which gets the optimized bases from the various methods as input. Thus
the simulation time do not have to coincide with the length of the POD
simulations. Further we have chosen a different random seed for statistical
initial conditions unless otherwise stated.
\subsection{Linear Diffusion}
For this problem the dynamics is given by Eq.(\ref{diffu1}).
\begin{figure}
\includegraphics[width=0.475\textwidth]{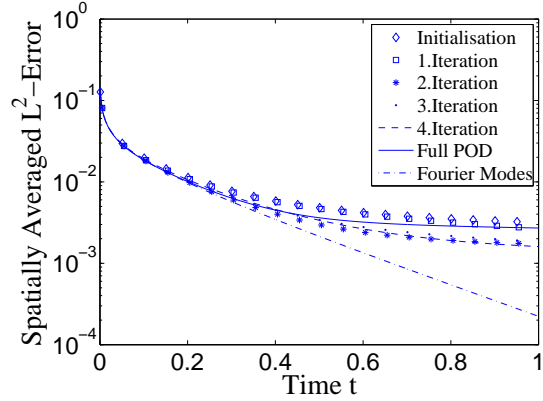}
\caption[Reduced diffusion equation $L^2$-Error]
{Reduced diffusion equation $L^2$-Error $E(t)$ for the analytical
 reduction (Fourier modes) the full
POD and DMRG POD after initialization and several iterations,
statistical initial condition,
N=40, $h_t=0.001$ and $d=0.05$. The error is expressed in units of $\Phi$,
for the time axis arbitrary units are employed. Note that for clarity not
all data points are shown as symbols.}
\label{fig:diffu_err1}
\end{figure}
\begin{figure}
\includegraphics[width=0.475\textwidth]{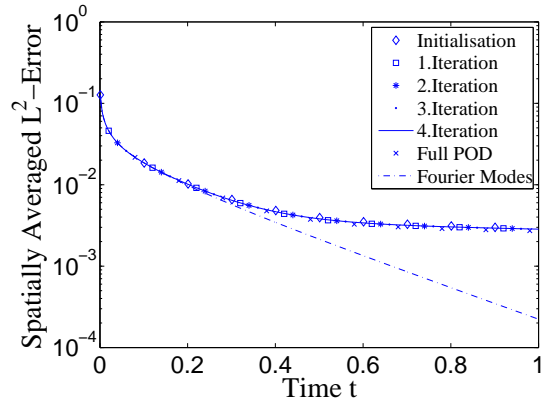}
\caption[Reduced diffusion equation $L^2$-Error, identical initialization]
{Reduced diffusion equation $L^2$-Error $E(t)$,
  identical statistical initialization.  The error is expressed in units
  of $\Phi$,  for the time axis arbitrary units are employed. Note that
  for clarity not all data points are shown as symbols.}
\label{fig:diffu_err2}
\end{figure}
The only nonzero contribution according to Eq.(\ref{system})
is $L\equiv\Delta_N$.
The eigenstates of $L$ are the sine/cosine or Fourier
modes whose contributions decay
over time with characteristic life-time inversely proportional to the
frequency/energy.
Standard DMRG can be viewed as an approximate diagonalization method for an
linear operator. Therefore it is very effective to find the optimal reduction
determined by the eigenstates, see Appendix~\ref{Appendix}, in the linear case.
In contrast to the diagonalization, POD as well as our method depends on
the initial conditions for the sample trajectories over which the averaging is
carried out. Both POD approaches cannot exploit the linearity of the evolution
equation. This affects the quality of the results for linear problems compared
to diagonalization-based methods.
Nevertheless, restriction to a few sample trajectories can also be an
advantage, since sometimes the interest lies on a certain region in phase
space. However, for the diffusion equation we choose normally distributed
initial conditions, i.e. the field $\Phi_0(x_i)$ is normally distributed.
This is then also true for the Fourier modes. By this choice
effectively the whole phase space will be sampled for a high enough number
of realizations. This is also due to the invariance of Eq.(\ref{diffu1})
under multiplication with a constant factor.

For the POD it is
important to integrate over long enough times. For short times the
state moves in the direction of the highest frequency modes which are
decaying most rapidly. Thus POD would give the wrong relevant modes.
The POD is in fact not a very appropriate tool to reduce the whole
phase space of the diffusion equation. In Fig.~\ref{fig:diffu_err1}
the error of the reduced fields $\hat{\Phi}$ is plotted in dependence of time.
There the time step was $dt=10^{-3}$ and the diffusion constant $d=0.05$.
The spatial resolution was $40$ lattice sites within the interval $[0,1]$.
In each POD step as well as for the error calculation the ensemble average,
compare with Eq.(\ref{correl}), has been
averaged over $50$ realizations of the initial conditions.
From this result we can state several things. First, all POD-based
methods show a remaining error in the long time limit. Second,
the initialization steps of DMRG POD gives already reasonable
results. An improvement due to the iteration is present, too.
Third, our algorithm is able to compute
the optimal reduction with even higher accuracy than the full POD.
The last point is only paradox on the first glance.
The inaccuracy of the full POD is in this case influenced from the
statistical initial conditions, in order to sample the full phase space.
Within our algorithm, much more initial conditions are taken into account as
the superblock POD is performed repeatedly.
This leads to a better
statistics. In Fig.~\ref{fig:diffu_err2} we have shown the same results
but using always the same initialization for calculating all PODs
(but of course not for the error calculation). It is clear that in this case,
our method has no advantage over the full POD anymore. On the other hand, the
results from our algorithm are not worse than that from the full
system POD, which is not clear a priori.
\begin{figure}
\includegraphics[width=0.475\textwidth]{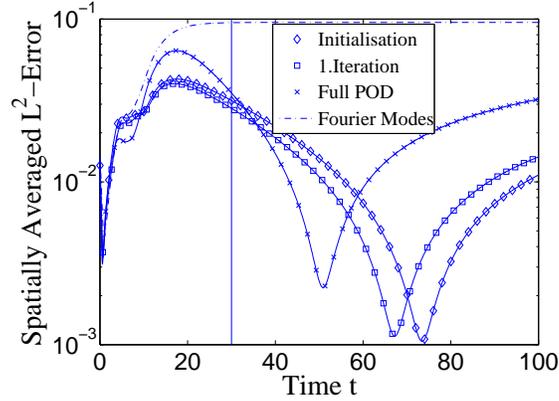}
\caption[Reduced Burgers equation $L^2$-Error]
{Reduced Burgers equation $L^2$-Error $E(t)$, deterministic initial condition,
  N=40, $h_t=0.02$, $d=0.01$ and $\nu=0.1$. The error is expressed in
  units of $\Phi$, for the time axis arbitrary units are employed. Note that
  for clarity not all data points are shown as symbols.}
\label{fig:burgers_err1}
\end{figure}
\begin{figure}
\includegraphics[width=0.475\textwidth]{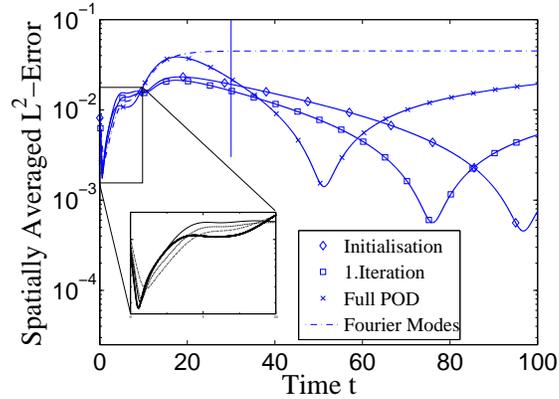}
\caption[Reduced Burgers equation $L^2$-Error]
{Reduced Burgers equation $L^2$-Error $E(t)$, deterministic initial condition,
  N=100, $h_t=0.005$, $d=0.01$ and $\nu=0.1$. The inset shows the begin of
  the error evolution enlarged. The error is expressed in units of $\Phi$,
  for the time axis arbitrary units are employed. Note that
  for clarity not all data points are shown as symbols.}
\label{fig:burgers_err2}
\end{figure}
\subsection{Burgers Equation}
The Burgers equation is given by Eq.(\ref{burgers1}). The discretization used
here is already described above.
We begin our analysis with the choice of deterministic initial condition for
the calculation of all PODs. In particular it is of the form
\begin{equation}\label{init1}
\Phi(t=0,x_i)= e^{-50(x_i-1)^2}\quad x_i=0\ldots 1.
\end{equation}
Fig.~\ref{fig:burgers_err1} and~\ref{fig:burgers_err2} show the results for
the $L^2$-Error of the evolution. Here we have used two spatial
resolutions, i.e. $N=40$ and $N=100$ nodes. The results are very similar.
In contrast to the previous calculations the simulation runs for the error
calculation are longer than the POD runs. The vertical line indicates the
time interval of the POD runs. Here we have to state that the Fourier mode
reduction is not optimal, which is not surprising as we consider a
nonlinear system and a very particular region of phase space.
Further, we see that the error
curves show a very pronounced minimum after which the approximation
seemingly breaks down. The corresponding time point lies well after
the POD time-span. These minima correspond to the fact that after the 
passing of the wavefront the profile becomes flat. The approximations
do not reproduce the average value accurately, but show a spurious drift.
The passing of the reduced (flat) states by the original (flat) state
creates the minima in Fig.~\ref{fig:burgers_err1}.

It is remarkable that our methods yield
better results than the POD within the POD time, even for the initialization.
Here it should be recalled that the POD is optimal only for
reconstructing the states used in the calculation.
As stated above, the reconstruction of the dynamics that created
these states, is a different thing as can be directly seen
from our results.
 
We continue our analysis of the Burgers equation by considering statistical
initial conditions. In contrast to the calculations for the diffusion equation
we have only three randomly sampled parameters in the initial condition.
It is given by a peak of various height $H$, width $W$ and position $X$. In
particular it is defined by the following equation
\begin{equation}\label{init_burgers1}
\Phi(t=0,x_i)=He^{-50W(x_i-X)^2}.
\end{equation}
Here, $H$ and $W$ are normally distributed whereas $X$ is uniformly
distributed.

\begin{figure}
\includegraphics[width=0.475\textwidth]{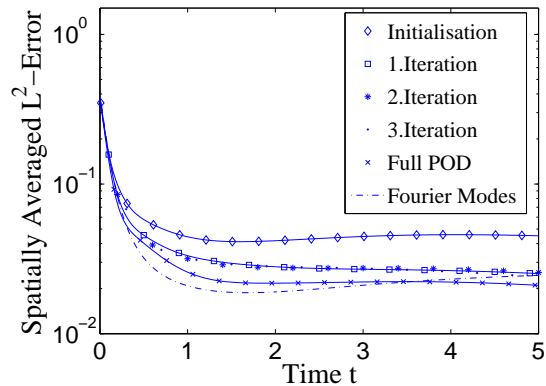}
\caption[Reduced Burgers equation $L^2$-Error]
{Reduced Burgers equation $L^2$-Error $E(t)$, statistical initial condition,
  N=20, $h_t=0.01$, $d=0.05$ and $\nu=0.1$. The error is expressed in
  units of $\Phi$, for the time axis arbitrary units are employed. Note that
  for clarity not all data points are shown as symbols.}
\label{fig:burgers_err3}
\end{figure}
The results are shown in Fig.~\ref{fig:burgers_err3}. For all methods
the error reaches a plateau very quickly. The performance of
the full system POD is slightly better than that of the DMRG POD.
However, the errors from our approach are of the same order as
from the full POD and one magnitude better than that of the Fourier mode
based reduction. Also the iteration brings an improvement which reaches
saturation already after the first step.
\begin{figure}
\includegraphics[width=0.475\textwidth]{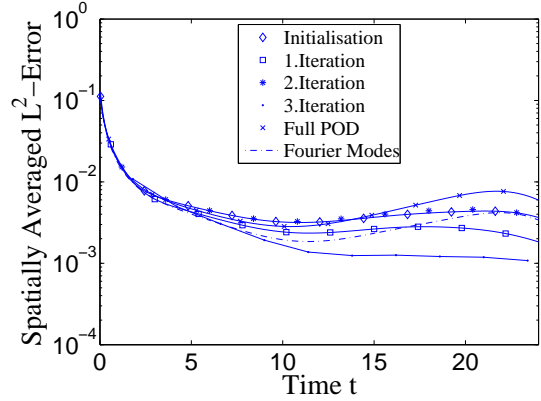}
\caption[Reduced nonlinear diffusion equation $L^2$-Error]
{Reduced nonlinear diffusion equation $L^2$-Error $E(t)$,
  statistical initial condition,
  N=30, $h_t=0.03$, $d=0.01$ and $a=0.5$. The error is expressed in
  units of $\Phi$, for the time axis arbitrary units are employed. Note that
  for clarity not all data points are shown as symbols.}
\label{fig:FN_err1}
\end{figure}
\begin{figure}
\includegraphics[width=0.475\textwidth]{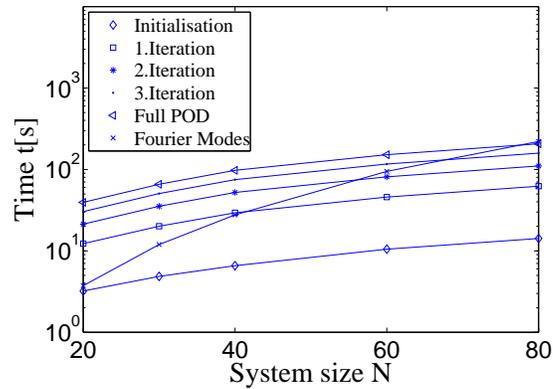}
\caption[Computing time for various system sizes and approaches] 
{Computing time for various system sizes and approaches, obtained by the
  reduced Burgers equation, statistical initial condition,
  N=40, $h_t=0.005$, $d=0.01$ and $\nu=0.1$.}
\label{fig:times1}
\end{figure}

For deterministic initial conditions the evolution of the error is not
monotonic in contrast to the case of statistical initial conditions.
This is due to the fact
that deterministic initial conditions can be considered more effectively
by the POD. The statistical initial conditions were drawn from a three, see
Eq.(\ref{init_burgers1}) or two, see Eq.(\ref{init_FN1}),
dimensional subspace which is reproduced poorly by a reduction
to a 4 dimensional space, which has to consider the time evolution also.

\subsection{Nonlinear Diffusion}
This system is defined by Eq.(\ref{FN1}).
As initial conditions we have chosen
a front with uniformly distributed position $X$ and normally distributed
height $H$:
\begin{equation}\label{init_FN1}
\Phi(t=0,x_i)=\frac{H}{2}\tanh((x_i-X)10).
\end{equation}
Under this conditions all methods were able to reproduce the dynamics well,
see Fig.~\ref{fig:FN_err1}.
Surprisingly the full POD method gave poorer results than even the
Fourier mode based reduction. This is to a lower extent also true for the
initialization run of the DMRG POD. The iteration lead to an improvement
although the 2nd iteration gave similar results as the initialization.
Further iterations again increase the accuracy, so no general statement
can be made.
After a fast saturation by applying the iteration procedure we observed
repeatedly a decay in the quality of the results, which we attribute
to the accumulation of numerical errors.

\section{Discussion}
\subsection{Computational Load}\label{comp_load}
For all calculation steps, e.g. diagonalization, Gram-Schmidt
orthonormalization etc., standard algorithms were
applied~\cite{golub, numerical_recipes}.
The focus was more on a concise assessment of the new algorithm instead of
an optimal solution of the toy problems. For the diagonalization of the
covariance matrix, e.g. first a
Householder-tri-diagonalization was performed~\cite{golub}, which is
an ${\mathcal O}(N^3)$ algorithm.
The calculation of the POD, either for the complete system or for the
superblock system was performed with the same routine. This comprised the
simulation as well as the diagonalization. 

For a POD the simulation of the
system in the time-span of interest is additionally necessary.
The required computational load for this simulation is implementation
dependent and is denoted with $S(\cdot)$.
Within our approach the simulation and diagonalization is performed
only on the superblock system.
Comparing the results from Fig.~\ref{fig:burgers_err1}
and ~\ref{fig:burgers_err2} suggests, that the
necessary number of iterations (sweeps) does not depend on the
full system size $N$.
If we denote the superblock size with $M$ and the
number of iterations with $N_i$ a naive estimation of the
computational load is given in table~\ref{load1}.
\begin{table}\label{load1}
\begin{tabular}{c|c}
full system POD & DMRG POD\\\hline
$O(N^3)+S(N)$ & $NN_iO(M) + S(M)$
\end{tabular}
\caption{Naive estimation of the computational load with full system size $N$,
superblock size $M$ and numbers of iterations $N_i$. The computational load
for the simulation is denoted with $S(\cdot)$}
\end{table}

For a more quantitative analysis we have measured the time necessary
to perform a full POD comprised of simulation and diagonalization.
Then we did the same for the initialization of the DMRG POD algorithm
including all simulation and diagonalization steps until the superblock
system described the full system of dimensionality $N$,
compare Fig.~\ref{fig:dmrg_init}, and a first reduced basis had been
calculated.
We also measured the computing time for one further iteration step
in the same way as for the initialization. The computing time is 
constant for all iteration steps so further data was extrapolated.
The underlying equation was the deterministic initialized Burgers
equation although the choice for an equation affects the computational
load only marginally.
As parameters we have chosen $h_t=0.005$, $d=0.01$ and $\nu=0.1$.
Fig.~\ref{fig:times1} shows a logarithmic plot of the results.
The DMRG POD approach shows a lower amount of computer time for the
initialization step. For higher system size this holds also for the
iterations. Generally the scaling with $N$ is favorable.
Note, that here only the DMRG method should be assessed. For this purpose
public assessable standard algorithms are sufficient, although
much more effective methods could be possible.
All calculations were performed on an Intel Dual Core machine, using
a single CPU.
\subsection{Stability}
\begin{figure}
\includegraphics[width=0.475\textwidth]{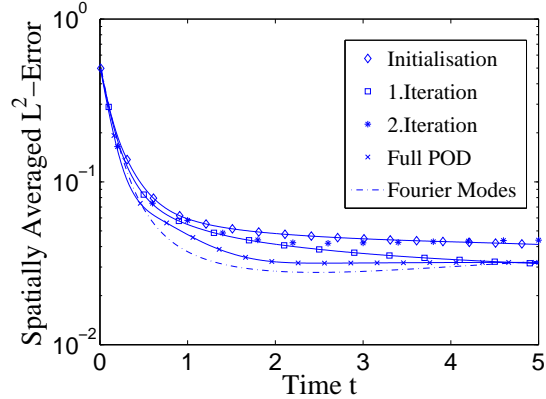}
\caption[Reduced Burgers equation $L^2$-Error]
{Reduced Burgers equation $L^2$-Error $E(t)$, statistical initial condition,
  N=40, $h_t=0.00625$, $d=0.05$ and $\nu=0.1$.
  The error is expressed in
  units of $\Phi$, for the time axis arbitrary units are employed. Note that
  for clarity not all data points are shown as symbols.}
\label{fig:burgers_err4}
\end{figure}

Many numerical schemes and the explicit Euler method in particular show
instabilities for certain parameter ranges. For the explicit Euler method
the stability condition is
\begin{equation}
  |1+\lambda h_t|\leq 1,
\end{equation}
where $\lambda$ is the largest eigenvalue of the generator of evolution
and $h_t$ the size of the time step. For the Laplace operator
the highest frequency component is the first to become instable while
increasing $h_t$. The eigenvalue is
$\frac{-4}{\delta x^2}\sin^2\left(\frac{\pi (N-1)}{N}\right)
\approx\frac{-4}{(\delta x)^2}$.
Consequently, we should have $h_t\leq\frac{(\delta x)^2}{2d}$,
$d$ being the diffusion constant. We have performed calculation directly
at this limit, see Fig.~\ref{fig:burgers_err4} and have seen no signs
of instability. Although the nonlinear Burgers equation was considered,
the previous point holds, since the symmetric derivative operator
has no nonzero real part eigenvalues.
By increasing $h_t$ the instability appears for all methods.
These calculations were only first tests and further work is required.
However, it can be assumed that in the DMRG POD calculation the instability 
originates only from the newly inserted nodes which correspond to the highest
spatial resolution. Combined implicit-explicit methods~\cite{fryxell_1986}
could be used to solve this problem. 
\begin{figure}
\includegraphics[width=0.475\textwidth]{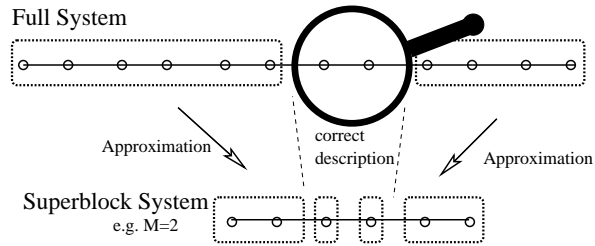}
\caption[]
{Pictorial representation of the approximation.}
\label{fig:lupe}
\end{figure}

\subsection{Interpretation of the Algorithm}
The various steps, necessary for a DMRG version of the Proper Orthogonal
Decomposition, seem to be complex on the first glance. It is also not clear
why this approach should be effective. We now shortly depict the basic idea
behind this algorithm. 

The DMRG algorithm decomposes the spatial domain but considers an interaction
of the domains due to the superblock concept. The single blocks, together
with information about interaction with neighboring blocks and the
reduced operators, describe spatial regions with a higher number of nodes
than the number of dofs actually retained in the block.
Inserting nodes described by
the full (yet already discretized) dynamics corresponds to
increasing the spatial resolution locally. The surrounding
blocks simulate the environment for a small subsystem with correct dynamics.
In the DMRG POD algorithm now the area with high resolution is moved through
the system. Thereby the parameters of superblock system, i.e. the
block basis and operator matrix elements, are adapted to approximate
the full system. For more graphical illustration see Fig.~\ref{fig:lupe}.
These arguments are a bit heuristic, but until now no rigorous proof
for the algorithm has be given. 
\subsection{Conclusion and Outlook}
To summarize, we have given a demonstration of applicability for a
new algorithm to calculate an approximate POD without
ever simulating the full system.
Our approach also makes practically no assumption on
the equations that define the dynamics.
The approach has been tested for linear systems where its performance
was even higher than the full system POD results but considerable
worse than the optimal reduction.
Several nonlinear systems have been considered. For the Burgers equation
the results of the full POD and our algorithm were comparable and
significantly better than a Fourier mode based reduction.

Further work on this topic will include extensions to higher dimensional
systems. A method for 2 and 3 dimensions is currently in progress.
Further, driven systems and systems with noise shall be implemented.
A closer analysis of the quality of the approximations as well as the
limitations of the method has to be performed.
We have access to the amount of discarded information from the
discarded eigenvalues in the truncations as well as in the POD steps. This
suggests an adaptive approach for the reduction.

I would like to thank Prof.F.~Schmid, Prof.~Ph.Blanchard and
Javier~Rodriguez-Laguna~\cite{our_project} for discussion and the
German science foundation (DFG) for support (DE 896/1-(1,2) 2004-2007).

\section{Appendix}\label{Appendix}
For completeness, we assess in the following the error of the
reduced evolution for the linear case. For the optimal reduction we require
a minimal $L^2$-error for the reduced field with respect to the un-reduced
evolution.
The full time evolution in the $N$ dimensional phase space is generated
by $L$ as
\begin{equation}\label{eq:dyn}
\Phi(t)=e^{(t-t_0)L}\Phi(t_0).
\end{equation}
The explicit Euler algorithm approximates this by
\begin{equation}
\Phi(t)\approx
\left(\openone+h_t L \right)\Phi(t_0).
\end{equation}
We assume that all eigenvalues of $L$ are negative or zero. A positive
eigenvalue would lead to an unbounded exponential growth in Eq.(\ref{eq:dyn})
which is unphysical.
Considering only linear projections the reduction is defined by
the operator $P$ which is the orthogonal projection to the relevant subspace
$\mbox{Range}(P)$. $P$ can be constructed from an ONB of this
space. Equivalently, it can be defined via 
the ONB (namely $C$) of $\mbox{Kern}(P)$ so that $P=\openone-CC^\dag$.

The reduced time evolution becomes
\begin{equation}
\hat\Phi(t) =e^{(t-t_0)PLP}P\Phi(t_0) = e^{(t-t_0)\hat{L}} \hat\Phi(t_0),
\end{equation}
since after each (infinitesimal) time step the component within the
irrelevant subspace, i.e. $\mbox{Kern}(P)$ are projected out.
For a general $P$ the eigenvectors of
$\hat{L}$ are not the same as for $L$, but known eigenvectors
of $\hat{L}$ are always the column vectors of $C$.
\subsection{Long Time Optimised Projection}
If we assume that the eigenvalues of $L$ are $\leq 0$,
for long times $t\gg 1$ the time evolution operators $e^{t\hat{L}},e^{tL}$
become the projectors onto the kernels of $L$ or $\hat{L}$, respectively.
In the eigenbasis $\psi^{eig}_i$ it is simply
\begin{equation}
\psi^{eig}_ie^{tL}\psi^{eig}_j =\delta_{ij}e^{t\lambda_i}
\end{equation}
The product of the reduced evolution operator $e^{tPLP}$ and $P$ converges
for long times to the projector onto $\mbox{Kern}(PLP)\cap\mbox{Range}(P)$.
More explicitly this is
\begin{eqnarray}
\lim\limits_{t\rightarrow\infty}e^{tL}&=&\left.\openone\right|_{\mbox{\small Kern}(L)},\\
\lim\limits_{t\rightarrow\infty}e^{tPLP}&=&\left.\openone\right|_{\mbox{\small Kern}(PLP)}.
\end{eqnarray}
This gives for the error
\begin{eqnarray}
E_\infty =\lim\limits_{t\rightarrow\infty}E\Phi(t)
=\left(
\left.\openone\right|_{\mbox{\small Kern}(L)}-
\left.\openone\right|_{\mbox{\small Kern}(PLP)}P
\right)\Phi(t_0)\label{longtimeE}.
\end{eqnarray}
In the long time limit we can obtain a zero error for all initial
conditions if we have
\begin{eqnarray}
\mbox{Kern}(PLP)\cap\mbox{Range}(P) \equiv \mbox{Kern}(L)\label{req1}.
\end{eqnarray}
This is achieved by requiring
\begin{eqnarray}
\mbox{Kern}(L) \subset \mbox{Range}(P),
&\quad\mbox{ and}\label{cond1}\\
\mbox{Range}(P) \quad L\mbox{ invariant.}\label{cond2}
\end{eqnarray}
as we show now.

Consider a $\phi\in\mbox{Range}(P)$. Then $P\phi=\phi$ and due to the
$L$-invariance of $\mbox{Range}(P)$ it is $L\phi\in\mbox{Range}(P)$
resulting in $PLP\phi = PL\phi = L\phi$. This gives for $P$ with
$\mbox{Range}(P)$ being $L$-invariant
\begin{eqnarray}
\mbox{Kern}(PLP)\cap\mbox{Range}(P) = \mbox{Kern}(L)\cap\mbox{Range}(P).
\label{invar1}
\end{eqnarray}
Eq.(\ref{req1}) can be retrieved from Eq.(\ref{invar1}) just by requiring
condition~(\ref{cond1}). Thus, in the long time limit Eq.(\ref{longtimeE})
becomes identically zero.

\subsection{Short Time Optimised Projection}
For short times we consider here the reduction from a $N$-dimensional to a
$(N-1)$-dimensional system.
For further reductions the results can be applied by iteration.
The projector $P$ becomes then $P_{ij}=\openone_{ij}-c_ic_j$ where $\vec{c}$
is the removed state.
In order to minimize the error for the short time evolution measured by the
$L^2$-norm we have to minimize
\begin{eqnarray}
E_s(t)&=&||e^{tL}\phi-e^{PLP}P\phi||_{2}\label{short-err1}\\
&\approx&||\left(\openone+tL-P-PLP\right)\phi||_{2}=||E\phi||_{2}.
\nonumber
\end{eqnarray}
Here we have already used an expansion in powers of $t$ and truncated after
the first order terms.

Since we have no information on $\phi$, we minimize Eq.(\ref{short-err1})
by using the Frobenius norm $|\cdot|_F$ of the error operator $E$.
The Frobenius norm is consistent with the $L^2$-norm~\cite{golub}, i.e.
\begin{eqnarray}\label{consist1}
||Ax||_{2}\leq|A|_F||x||_{2} \quad\forall A\in R^{n\times n},x\in R^n.
\end{eqnarray}
By inserting $P=\openone-C$ we get for the error operator
\begin{eqnarray}
E=&\openone-P+t(L-(\openone-C)L(\openone-C))\nonumber\\
=&C+t(L-L+LC+CL-CLC)\nonumber\\
=&C+t(LC+CL-CLC).
\label{consist2}
\end{eqnarray}
We assume $L$ to be symmetric, i.e. $L_{ij}=L_{ji}$.
Thus $L$ has an orthonormal eigenbasis
$\left\{\varphi_{i\alpha}\right\}_{\alpha=1\ldots N}$ where the columns are the
eigenvectors of $L$. The eigenvalues are $\lambda_\alpha$ and the matrix
elements of the error operator $E$ are decomposed in this basis as
\begin{eqnarray}
E_{ij} &=& \sum\limits_{\alpha\beta}
\varphi_{\alpha i} E_{\alpha\beta} \varphi_{\beta i}\nonumber\\
&=&C_{ij}+t\sum\limits_{n}\left(L_{in}C_{nj}+C_{in}L_{nj}
-\sum\limits_{m}C_{in}L_{nm}C_{mj}\right)\label{error}
\end{eqnarray}
with
\begin{eqnarray}
C_{ij} &=&\sum\limits_{\alpha\beta}
\varphi_{\alpha i}c_\alpha c_\beta \varphi_{\beta j},\label{Cij}\\
\sum\limits_{mn}C_{in}L_{nm}C_{mj} &=&
\sum\limits_{\alpha\beta nm}
\varphi_{\alpha i}c_\alpha c_nL_{nm}
c_m c_\delta\varphi_{\beta j},\\
\sum\limits_{n}L_{in}C_{nj} &=&
\sum\limits_{\alpha\beta n}
\varphi_{\alpha n}L_{i n}c_\alpha c_\beta\varphi_{\beta j},\\
\sum\limits_{n}C_{in}L_{nj} &=&
\sum\limits_{\alpha\beta n}
\varphi_{\alpha i}c_\alpha c_\beta L_{nj}\varphi_{\beta n}.\label{CLC}
\end{eqnarray}
We use the orthogonality of the $\varphi_\alpha$, i.e.
\begin{equation}
\sum\limits_{i}\varphi_{\alpha i}\varphi_{\beta i}=\delta_{\alpha\beta}=
\sum\limits_{i}\varphi_{i\alpha}\varphi_{i\beta}.
\end{equation}
In the eigenbasis the removed degree of freedom $\vec{c}$ can be written as
$\vec{\tilde{c}}$ with components
\begin{eqnarray}
\tilde{c}_i=\sum\limits_{\alpha}\varphi_{\alpha i} c_\alpha
\quad,\quad
c_\beta=
\sum\limits_{i\alpha}\varphi_{\alpha i}\varphi_{\beta i}
c_\alpha
=
\sum\limits_{i}\varphi_{\beta i} \tilde{c}_i.
\end{eqnarray}

The average of $L$ in the removed state $\vec{c}$, is
\begin{eqnarray}
\left<L\right>_{\vec{c}}&:=&
\sum\limits_{nm}c_nL_{nm}c_m
=\sum\limits_{nmij}
\varphi_{ni}\tilde{c}_iL_{nm}\varphi_{mj}\tilde{c}_j\\
&=&\sum\limits_{nij}
\varphi_{ni}\tilde{c}_i\lambda_{j}\varphi_{nj}\tilde{c}_j
=\sum\limits_{i}
\tilde{c}_i^2\lambda_i\nonumber
\end{eqnarray}
The matrix elements from Eq.s(\ref{Cij}-\ref{CLC}) become
\begin{eqnarray}
C_{ij} &=&\tilde{c}_i \tilde{c}_j,\\
\sum\limits_{mn}C_{in}L_{nm}C_{mj}&=&
\tilde{c}_i \left<L\right>_{\vec{c}}\tilde{c}_j,\\
\sum\limits_{n}L_{in}C_{nj}&=&
\sum\limits_{\alpha n}
\varphi_{\alpha i}L_{\alpha n}c_n \tilde{c}_j\nonumber\\
&=&
\sum\limits_{n}
\lambda_i\varphi_{ni} c_n \tilde{c}_j\nonumber\\
&=&
\lambda_i\tilde{c}_i\tilde{c}_j,\\
\sum\limits_{n}C_{in}L_{nj}&=&
\sum\limits_{n}
\tilde{c}_i c_n\lambda_j\varphi_{nj}\nonumber\\
&=&\tilde{c}_i \tilde{c}_j\lambda_j.
\end{eqnarray}
Thus for the matrix elements of the error operator we obtain
\begin{eqnarray}
E_{ij} = \tilde{c}_i \tilde{c}_j\left(
1+t\left(\lambda_i + \lambda_j -\left<L\right>_{\vec{c}}\right)
\right).
\end{eqnarray}
We minimize the Frobenius norm of $E$ given by
\begin{eqnarray}
\label{err_frobenius}
|E|_F=\sum\limits_{ij}|E_{ij}|^2 = \sum\limits_{ij}
\tilde{c}_i^2 \tilde{c}_j^2\left(
1+t\left(\lambda_i + \lambda_j - \left<L\right>_{\vec{c}} \right)
\right)^2
\end{eqnarray}
for a normalized $\vec{c}$, i.e.
\begin{equation}
1=||c||_2^2=\sum\limits_ic_i^2=\sum\limits_i\tilde{c}_i^2.\label{constraint}
\end{equation}
Since $E$ is a linear operator it follows that
$||E{\bf x}||_2=||{\bf x}||_2||E\hat{\bf x}||_2$ with
${\bf x}=\hat{\bf x}||{\bf x}||_2$. Without no
restriction to $||{\bf x}||_2$ the zero vector would always minimize
$||E{\bf x}||_2$. 
Furthermore, each lower bound $K$ for
$||{\bf x}||_2$ will lead to the same
$\hat{\bf x}$ with $||\hat{\bf x}||_2=K$. This is not true for the general
nonlinear case as in~\cite{Degenhard_Rodriguez-Laguna_2002}.

Incorporating this condition $|E|_F$ reduces to
\begin{eqnarray}
|E|_F^2&=&
1+2t\left<L\right>_{\vec{c}}+2t\left<L\right>_{\vec{c}}
-2t\left<L\right>_{\vec{c}}
+t^2 \left<L\right>_{\vec{c}}^2 \nonumber\\
&+&2t^2 \left<L\right>_{\vec{c}}^2
-2t^2 \left<L\right>_{\vec{c}}^2 + t^2 \left<L\right>_{\vec{c}}^2
-2 t^2 \left<L\right>_{\vec{c}}^2 +t^2 \left<L\right>_{\vec{c}}^2\nonumber\\
&=&1+2t\left<L\right>_{\vec{c}}+t^2 \left<L\right>_{\vec{c}}^2
=\left(1+t\left<L\right>_{\vec{c}}\right)^2\nonumber\\
\Rightarrow
|E|_F&=&\left|1+t\left<L\right>_{\vec{c}}\right|.\label{err_frobenius2}
\end{eqnarray}
Consequently, in order to minimize $|E|_F$ we have to minimize
$\left<L\right>_{\vec{c}}$.

The minimization itself is performed using Lagrangian multipliers
for the constraint Eq.(\ref{constraint}). The
necessary condition for a minimum is
\begin{eqnarray}
0&=&\frac{\partial}{\partial \tilde{c}_k}
\left(
\left<L\right>_{\vec{c}}+\eta\left(1-||\vec{c}||_2^2\right)
\right)\nonumber\\
&=&
\frac{\partial}{\partial \tilde{c}_k}
\sum\limits_{i}
\left(
  \tilde{c}_i^2\lambda_i
  +\eta\left(1-\tilde{c}_i^2\right)
\right)\\
&=&2\tilde{c}_k\left(\lambda_k-\eta\right).\nonumber
\end{eqnarray}
This is true if either $\tilde{c}_k=0$ or $\eta=\lambda_{k}$. The last
equation can only be true for a single value of $\lambda_{k}$.
We denote the nonzero component as $\tilde{c}_{k^\prime}\neq0$ and
$\tilde{c}_k=\delta_{kk^\prime}\tilde{c}_{k^\prime}$. From
Eq.(\ref{constraint}) it follows further that
$\tilde{c}_k=\delta_{kk^\prime}$.

Inserting this in Eq.(\ref{err_frobenius2}) we obtain
\begin{eqnarray}
|E|_F=
\left|
1+t
\sum\limits_{k}\tilde{c}_k^2\lambda_k
\right|
=\left|
1+t\lambda_{k^{\prime}}
\right|.
\end{eqnarray}
For small $t$ , i.e. $t<|\lambda_i|^{-1}$ $\forall i$ this is clearly minimal
if we choose $\lambda_{k^{\prime}}$ to be the smallest eigenvalue.

Further iterations, e.g. $n$ times, of selecting the irrelevant states
remove successively  the eigenstates corresponding to the $n$ lowest
eigenvalues. This is due to the fact that the spaces
Kern($C$)~$\equiv$~Range($P$) and 
Range($C$)~$\equiv$~Kern($P$) are by construction $L$ invariant.
This also makes the iteration unambiguous, a feature that is in general
not present for nonlinear problems.

Note also that since $\lambda_i\leq0$ the reduced states always belong
to Range($L$) as long as any remaining eigenvalue, i.e. an eigenvalue 
of $P_{n-1}LP_{n-1}$ is nonzero. Here, $P_{n-1}$ results from the
previous reduction step. In this case the error always vanishes for long times.

Summarizing, the optimal short time projection leads to results that are
not only consistent with the long time accuracy requirements, but even
include them.

\bibliography{../nonlin_dyn}

\end{document}